\documentclass[final]{svjour3}
\usepackage[dvipdfmx]{graphicx}
\usepackage{rotating}
\usepackage{amssymb}
\usepackage[square, numbers, sort]{natbib}
\makeatletter
\journalname{Journal of Low Temperature Physics}
\usepackage{comment}
\usepackage{hyperref}
\usepackage[bottom]{footmisc}
\usepackage{siunitx} % allows the use of \num{6e24} and \SI{value}{unit commands}
\usepackage{mathtools} % for additional math commands over basic LaTeX 
\usepackage{xcolor} % to use \textcolor{color}{text}
\usepackage{textcomp} 
\usepackage{caption}
\usepackage{subcaption}

\begin{document}

\newcommand{\hdblarrow}{H\makebox[0.9ex][l]{$\downdownarrows$}-}
\title{Millimetre Wave Kinetic Inductance Parametric Amplification using Ridge Gap Waveguide}

\author{D. Banys \and M. A. McCulloch \and T. Sweetnam \and V. Gilles \and L. Piccirillo}

\institute{Jodrell Bank Centre for Astrophysics,  University of Manchester,\\ Manchester, M13 9PY, United Kingdom\\
Tel.: +44 161 306 6470\\ 
\email{danielius.banys@manchester.ac.uk}}

\maketitle

\begin{abstract}

We present the design and simulation methodology of a superconducting ridge-gap waveguide (RGWG) as a potential basis for mm-wave kinetic inductance travelling wave parametric amplifiers (KI-TWPAs). A superconducting RGWG was designed using Ansys HFSS to support a quasi-TEM mode of transmission over a bandwidth of 20 to 120 GHz with its internal dimensions optimised for integration with W-band rectangular waveguide. A design of an impedance loaded travelling wave structure incorporating periodic perturbations of the ridge was described. A method to simulate the nonlinear kinetic inductance via user-defined components in Keysight's ADS was outlined, which yielded the power dependent S-parameters and parametric signal gain. A RGWG with a \SI{30}{\nano\meter} NbTiN coating and \SI{5}{\micro\meter} conductor spacing, corresponding to a kinetic inductance fraction $\alpha \sim 60\%$ was used for the description of a KI-TWPA with 900 perturbations equivalent to a physical length \SI{25}{\centi\meter} that achieved more than \SI{10}{\decibel} of signal gain over a 75--110 GHz bandwidth via 4-wave mixing (4WM).


\end{abstract}

\section{Introduction}

Kinetic inductance travelling-wave parametric amplifiers (KI-TWPAs) promise high gain, large bandwidth and potential for quantum limited noise performance \cite{Esposito2021}. Currently, most KI-TWPAs \cite{Eom2012, Vissers2016, Malnou3WM} have been fabricated for sub-20 GHz and \si{\milli\kelvin} temperature operation. However, there is interest in extending this to higher frequency applications where quantum limited amplification is needed such as dark matter detection \cite{Beurthey2020madmax} and high frequency quantum computing \cite{Fermarzi2021}. At  mm-wave energy scales these superconducting circuits are less sensitive to thermal background noise due to the higher photon energies involved. Hence, these paramps may achieve their low noise figures at temperatures of \SI{1}{\kelvin}, which are accessible via higher heat lift fridges such as adsorption coolers. Additionally, even at temperatures of \SI{4}{\kelvin} KI-TWPAs are competitive with their transistor-based counterparts while producing an order of magnitude smaller heat load \cite{Malnou2021_4K}. Recently, narrowband parametric amplification has been achieved at \SI{30}{\giga\hertz} \cite{Banys2020} and at \SI{95}{\giga\hertz} \cite{Anferov2020}, while the highest reported frequency in a travelling wave design \cite{Shu2021} is approaching \SI{30}{\giga\hertz}.

At mm-wave frequencies KI-TWPAs based on conventional transmission lines such as co-planar waveguides and microstrips face challenges in connecting their \si{\micro\meter} size traces to external feeding structures without suffering from large dielectric losses. A potential method to alleviate dielectric losses at frequencies over 100 GHz is to use more exotic transmission lines that incorporate a hollow gap between conducting surfaces instead of one filled with dielectric and can be coupled to waveguide feeds via low loss transitions. One such structure is ridge gap waveguide (RGWG)\cite{Kildal2011} which is readily scalable to frequencies of a few hundred GHz \cite{Rahiminejad2012} and can support quasi-TEM mode of transmission over an octave of bandwidth. Coating these RGWGs with a thin superconducting (SC) film such as NbTiN and maintaining a conductor separation in the micrometer to sub-micrometer range can help achieve kinetic inductance fractions in excess of 90\%, which makes these transmission line structures potential candidates for mm-wave KI-TWPAs.

\section{High Kinetic Inductance Ridge Gap Waveguide}

\begin{figure}
    \centering  
    \begin{subfigure}[b]{0.49\textwidth}
        \centering
        \includegraphics[width=\textwidth]{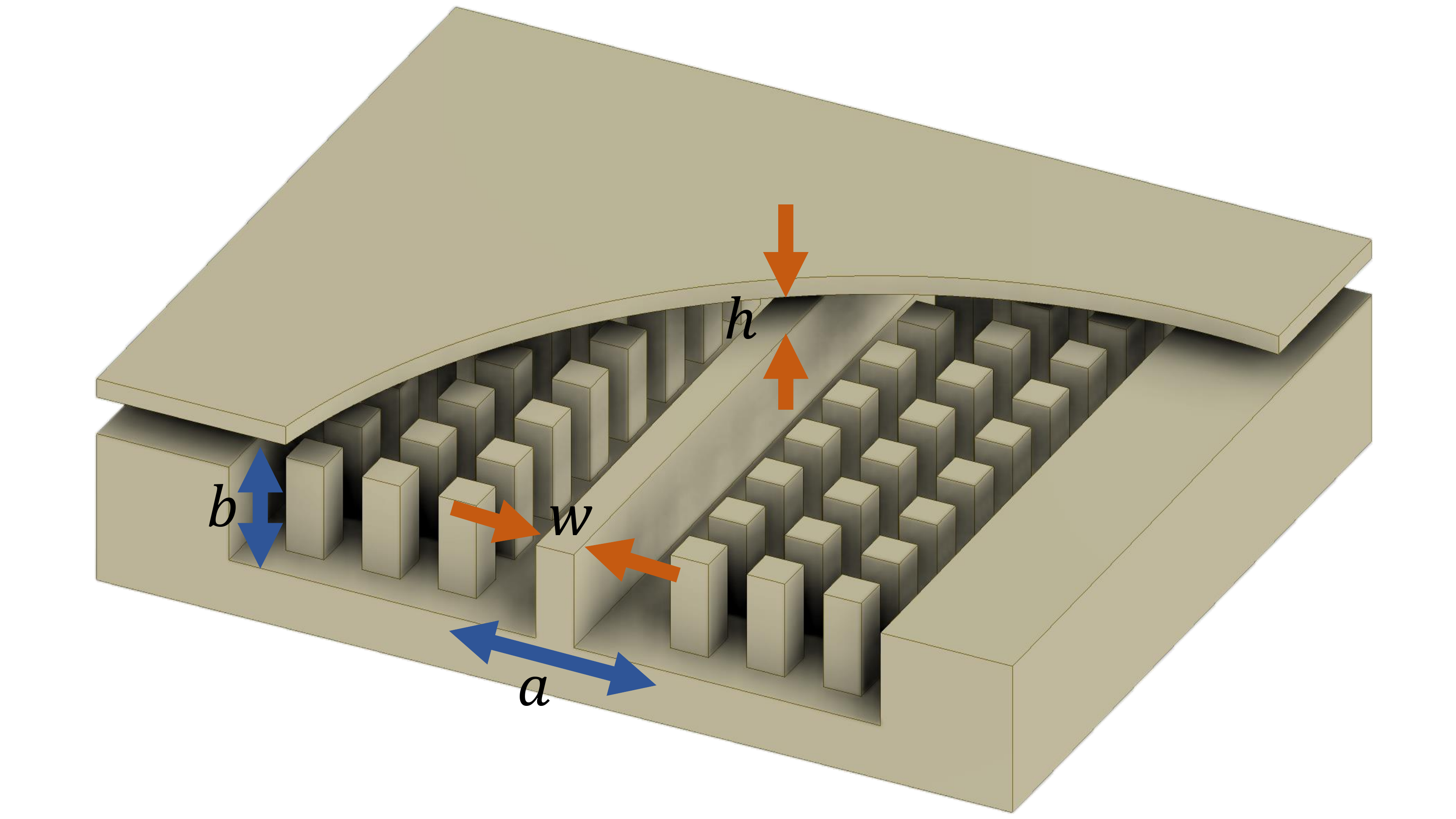}
    \end{subfigure}
    \hfill
    \begin{subfigure}[b]{0.49\textwidth}
        \centering
        \includegraphics[width=\textwidth]{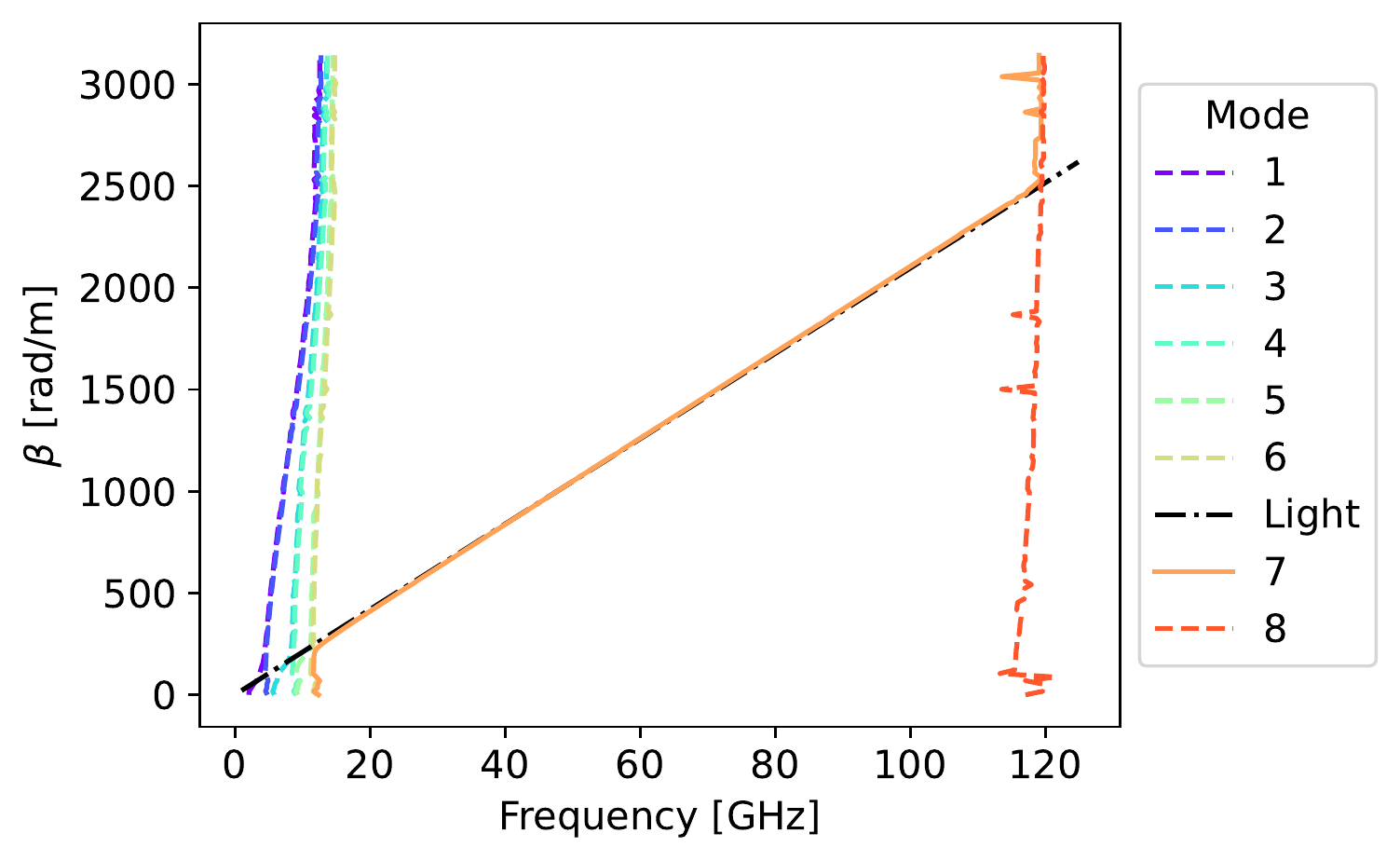}
    \end{subfigure}
	\caption{Left: A cross section view of RGWG model showing the central ridge of width $w=$ \SI{0.5}{\milli\meter} in a groove of width $a=$ \SI{2.54}{\milli\meter} and height $b=$ \SI{1.27}{\milli\meter}, with 3 rows of pins with width and separation of \SI{0.5}{\milli\meter}. The dimensions of $a$ and $b$ were chosen to match that of W band rectangular waveguide. Right: The dispersion of the first 8 modes from the eigenmode simulation of RGWG in HFSS. Modes 1-6, and 8 are the spurious modes that propagate in the whole internal volume and signify the lower and upper cut-off frequencies at which the bed of nails protects the propagation of the RGWG mode. As shown by the overlaid "light" line ($\beta = 2\pi f/c$) mode 7 is the quasi-TEM RGWG mode which is the sole propagating mode between 20 to 120 GHz (Color figure online).}
    \label{fig:ridgegapWG}	
\end{figure}

RGWG as shown in Fig.~\ref{fig:ridgegapWG} consists of two parallel conducting plates where one of these is textured with a wave-guiding ridge that is surrounded by a periodic pin surface. This surface, often referred to as a bed of nails, on its own is able to stop the propagation of parallel plate modes that may exist in the hollow enclosed cavity if the nails were absent. This protecting characteristic is exhibited over a certain bandwidth that depends on the length of the nails and the gap to the top surface \cite{Kildal2011}. A wave guiding channel may be created between these nails, where the nails act as soft magnetic boundaries at the edges of the channel, creating a type of groove gap waveguide (GGWG). Finally, adding a central ridge to this groove allows the propagation of a TEM mode with fields constrained to the volume between the top of the ridge and the opposing lid.

RGWG can be fabricated using conventional metal milling or silicon micro-machining techniques \cite{Rahiminejad2012}. In both cases a SC coating is necessary to provide a nonlinear kinetic inductance, which is required for parametric processes to take place. The extent to which this nonlinearity dominates over the total structure inductance is dictated by the kinetic inductance fraction, $\alpha$, which is dependent on the waveguide geometry and film properties. It is possible for $\alpha\rightarrow1$ to be achieved when the SC coating is thin enough such that $t\ll\lambda_L$, where $\lambda_L$ is the London penetration depth, and the hollow gap is small such that $h<1$ \si{\micro\meter}, which can be achieved via a deposition of a thin spacer along the outer perimeter of one of the involved bodies.

RGWG design begins with the engineering of the pin surface which creates the stopband for any global parallel plate or cavity modes that may exist. It was found that the pin surface protecting bandwidth was maximised when the nail width and separation were equal. A nail width of 0.5 mm was chosen with consideration to the mechanical fragility of the nails. Following this, the GGWG and RGWG are considered, which are dictated by the groove width, $a$, and ridge width $w$, respectively. The groove width was set to match the long wall of W band rectangular waveguide and three rows were chosen to provide sufficient isolation to the device edges in the GGWG region that would be needed in the transition between rectangular waveguide and RGWG via a taper. The GGWG mode fields decay by $\sim$ 30 dB per row of nails. The upper cut-off of the RGWG mode can be affected by $w$ especially when $w\rightarrow a$, since the added presence of a high KI coating can cause higher order RGWG modes to propagate. The main tool for designing and simulating these structures was Ansys HFSS\footnote{https://www.ansys.com/products/electronics/ansys-hfss}. The superconducting coating was included in S-parameter simulations via frequency dependent surface impedance boundaries \cite{Belitsky2006} that are imported from a calculation of the Zimmermann conductivity \cite{Zimmermann1991}. Specifically, the eigenmode solver of HFSS was used to calculate the dispersion properties of pin surface and RGWG, which is shown in Fig. \ref{fig:ridgegapWG}.

A thin SC coating enhances the linear kinetic inductance and sets the critical current of the line which needs to be reached for maximum non-linearity. Since the top of the ridge is easily acessible, the thin SC coating can be applied to this surface. A thin coating ($<\lambda_L$) on the textured surface deposited by a technique such as magnetron sputtering may cause current restrictions or discontinuities on sharp edges of the pins and ridges. These coating irregularities may give rise to undesired dissipative behaviour in regions of high current flow where the critical current may be surpassed locally. This may be alleviated by depositing a much thicker coating ($\gg\lambda_L$) on the textured surface or using a deposition technique such as atomic layer deposition, which can produce a thin conformal coating on all edges.

For this work, NbTiN coatings with a critical temperature $T_c \sim $ \SI{12.5}{\kelvin} and resistivity in the range of 200 \si{\micro\ohm} \si{\centi\meter} are considered. This coating would allow paramp operation at $\sim 1-4$ K with gain and noise figures comparable to equivalent transistor-based amplifiers \cite{Malnou2021_4K}. At these temperatures dissipation should be low due to the $\exp{(-T_c/T)}$ scaling of the SC surface resistance \cite{Gurevich2017}.

The linear kinetic inductance, $L_{k0}$, introduced via the SC coating scales with the dimensions and physical properties of the film. The kinetic inductance scales non-linearly with current \cite{Semenov2020} and when expanded to second order can be expressed as $L_k(I) = L_{k0}\left[1 + (I/I_*)^2\right]$, where $I_*$ is the scaling current which is dependent on the SC material properties and the geometry of the trace. 

For parametric processes to be explored it is first necessary to estimate the values of the low current KI and scaling current. The value of $L_{k0}$ per square can be estimated \cite{Belitsky2006} using $L_{k0}= \mu_0\lambda_L\tanh{(t/\lambda_L)}$. To determine the extent to which the KI non-linearity dominates over the total inductance the KI fraction can be calculated using, $\alpha = L_k/L_g+L_k$, where $L_g=\mu_0h$ is the geometric inductance for a RGWG assuming a parallel plate (separated by $h$) approximation.

The scaling current is sensitive to a number of factors \cite{Zolochevskii2014} including the substrate roughness and composition, film geometry, deposition conditions and operating temperature which means it can vary over orders of magnitude for a given SC material. Analytic solutions of the scaling current usually only set the upper bound for this value. The scaling current, which is of the order of the critical current, $I_c$, takes on different solutions based on the current distribution on the SC trace. For a conductor of thickness $t \ll \lambda_L$ and width $w \ll \lambda_\perp$, where $\lambda_\perp =2\lambda_L^2/t$ is the perpendicular penetration depth, the current density can be assumed to be constant across the film \cite{Pearl1964}. This low temperature scaling current can be estimated \cite{Zmuidzinas2012, Shu2021} using $I_*=wt\kappa_*\sqrt{N_0\Delta^2/\mu_0\lambda_L^2}$, where $\kappa_*$ is constant of order 1 \cite{Zhao2020} and $N_0$ is the number density at the Fermi level. However, if the trace is wider than the perpendicular penetration depth $w \gg \lambda_\perp$, the current distribution becomes non-uniform and accumulates at the edges of the film \cite{Zolochevskii2014}. In this work, the SC traces are in the 100s \si{\micro\meter} which means solutions of the wide film critical current should be explored. In the case of a pristine film with perfect edges the critical current \cite{Zolochevskii2014} in SI units is given by

\begin{equation}
    I_c(T) = \frac{2\Phi_0}{3\sqrt{3\pi}\xi(0)\mu_0}\left(\frac{w}{\lambda_\perp(0)}\right)^{1/2}(1-T/T_c),
    \label{eq:IcWide}
\end{equation}

\noindent where $\mu_0$ is the permeability of free space, $\Phi_0$ is the magnetic flux quantum and $\xi(0)$ is the zero temperature Ginzburg-Landau coherence length. The perfect edge stipulation is important since $I_c$ in a wide film is dependent on the pinning properties that arise at the edges of the trace \cite{Zolochevskii2014}. Therefore, Eq. \ref{eq:IcWide} provides only the upper limit for the wide film critical current. The scaling current should be measured experimentally by measuring the kinetic inductance induced frequency shift of a resonant cavity with ac or dc current \cite{Eom2012}. 

\section{Travelling-Wave Paramp using Ridge Gap Waveguide}

In the presence of a strong pump tone and weak signal a four photon mixing process can take place \cite{Erickson2017}, referred to as four wave mixing (4WM). These tones are related by $2\omega_p = \omega_s + \omega_i$, where $\omega_{p,s,i}$ are the frequencies of the pump, signal and idler tones, respectively.

For sufficient parametric gain the RGWG needs to be extended into an electrically long structure spanning hundreds of wavelengths. In addition, the phase relationship between the pump, signal and idler tones should be considered such that the dispersion difference is $\Delta\beta = \beta_s+\beta_i-2\beta_p=0$, where $\beta_{p,s,i}$ are the propagation constants of the involved tones. The nonlinear SC line introduces a power dependent phase shift, which must be compensated for by introducing anomalous dispersion \cite{Eom2012, Bockstiegel2014, Vissers2016, Shan2016} in order to satisfy the condition for wideband exponentially scaling gain. If this condition is not met, the gain will scale quadratically with propagation length. 

The simplest dispersion engineering method involves the perturbation of the transmission line every $\lambda_{\text{per}}/2$ such that stopbands are created at integer values of frequency $f_{\text{per}}$ near which anomalous dispersion is introduced. By positioning the pump tone near these dispersion features, the nonlinear phase slippage may be partially or even fully cancelled out.

\begin{figure}
    \centering
    \begin{subfigure}[b]{0.49\textwidth}
        \centering
        \includegraphics[width=\textwidth]{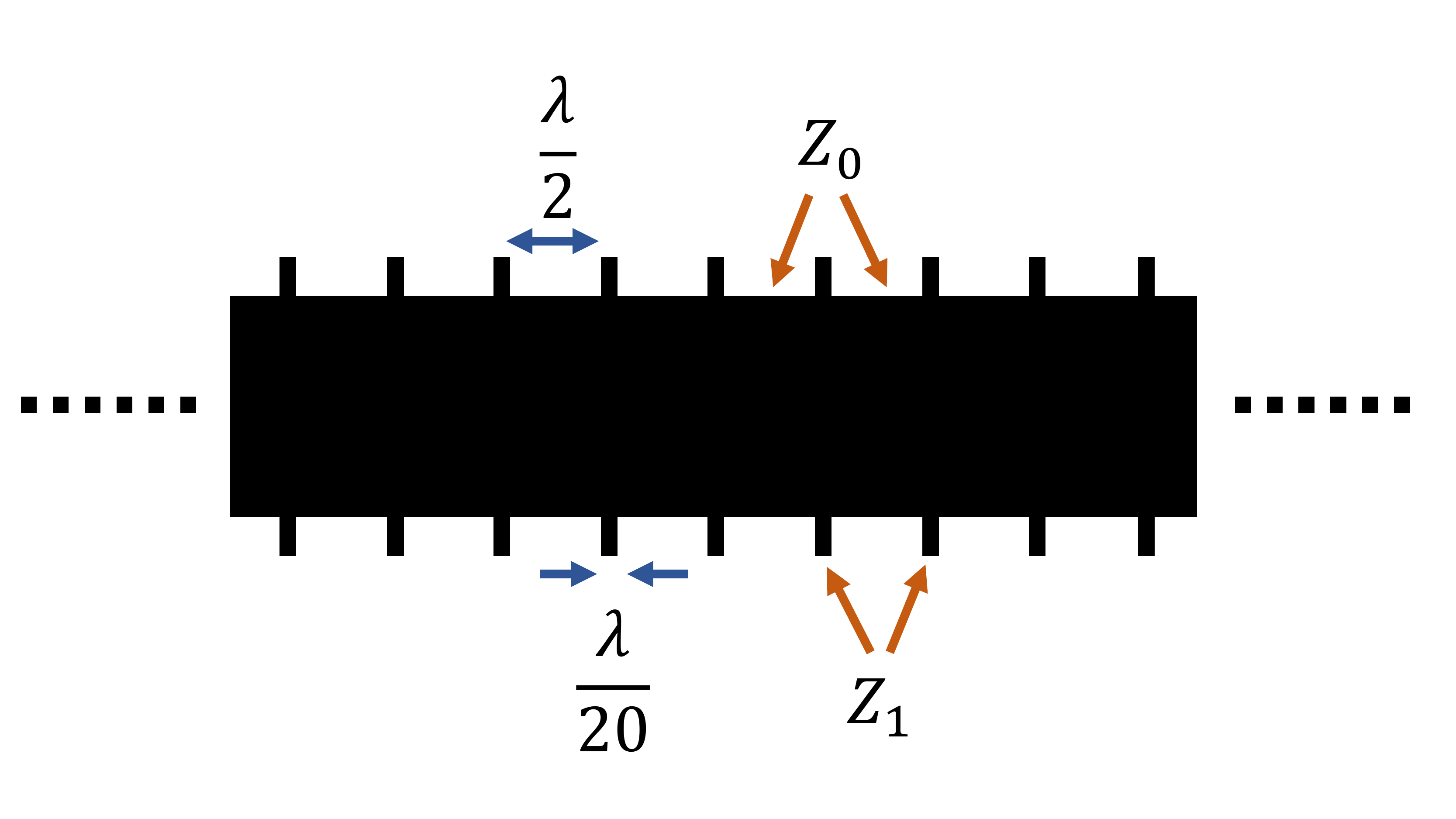}
    \end{subfigure}
    \hfill
    \begin{subfigure}[b]{0.49\textwidth}
        \centering
        \includegraphics[width=\textwidth]{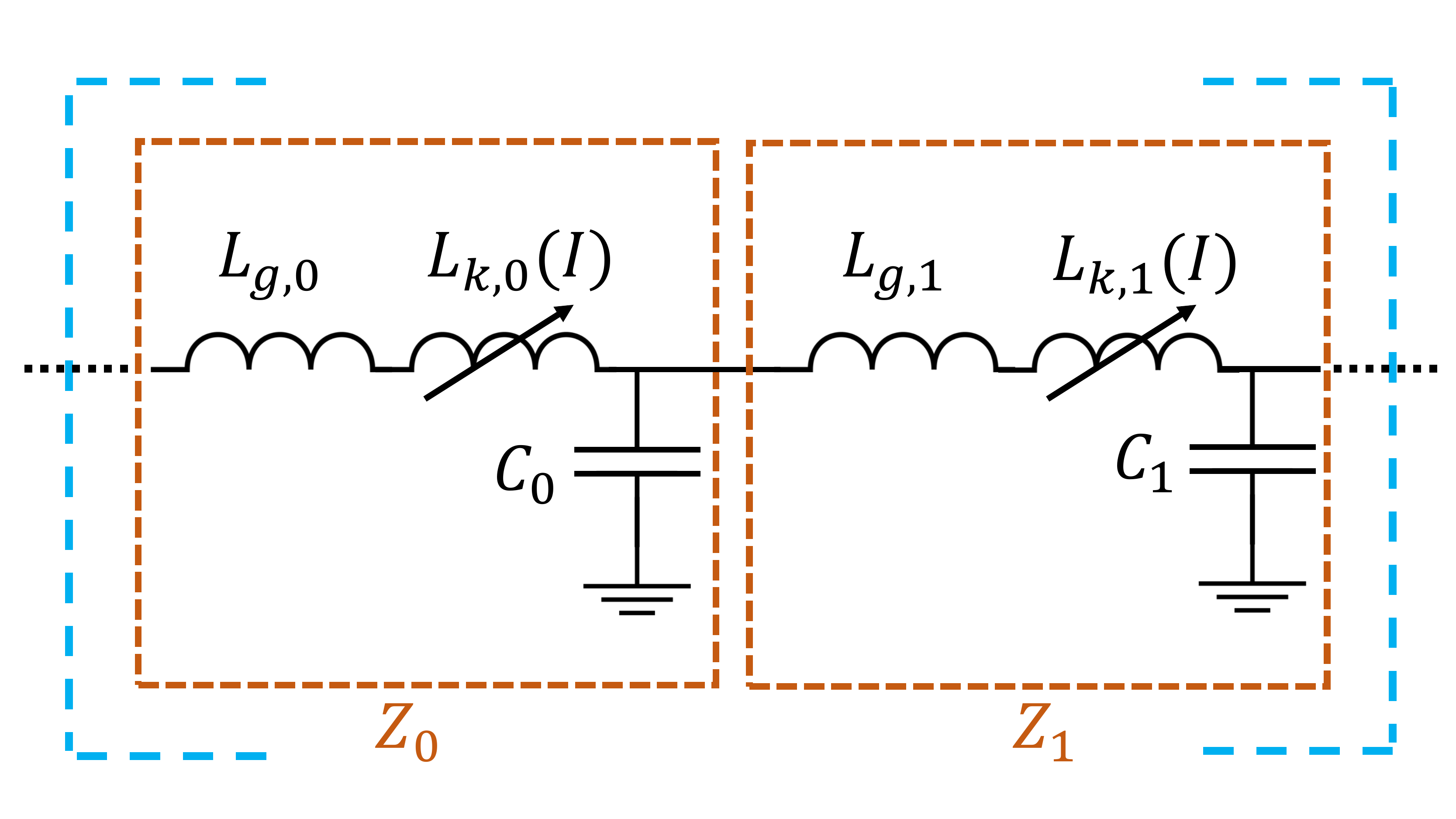}
    \end{subfigure}
	\caption{Left: A schematic of a section of the ridge in a dispersion engineered RGWG using design principle from \cite{Vissers2016}. The physical width of the main line is \SI{500}{\micro\meter}, while the repeating perturbations are \SI{650}{\micro\meter} wide, and $\lambda/20=$ \SI{25}{\micro\meter} long. The separation of loading midpoints is $\lambda/2=$ \SI{250}{\micro\meter}. Right: An equivalent circuit model of an unperturbed and perturbed section of the RGWG. The impedances of the main line and perturbed sections were chosen to be $Z_0 \sim$ \SI{6}{\ohm} and $Z_1 \sim $ \SI{4.75}{\ohm}, respectively, which were confirmed in HFSS simulations and included the kinetic inductance contribution of two surfaces coated with 30 nm of NbTiN ($\lambda_L \sim$ \SI{360}{\nano\meter}). Using such a low impedance increased the induced AC current for a given input power, which had the effect of reducing the necessary drive power to stimulate nonlinear behaviour. In addition, the chosen ridge to lid spacing of $h = $ \SI{5}{\micro\meter} resulted in $\alpha \sim$ \SI{60}{\percent}. The corresponding equivalent circuit parameters were $L_{g,0} \sim$ \SI{12.5}{\nano\henry\per\meter}, $L_{g,1} \sim$ \SI{9.5}{\nano\henry\per\meter}, $L_{k,0} \sim$ \SI{22}{\nano\henry\per\meter}, $L_{k,1} \sim$ \SI{17}{\nano\henry\per\meter}, $C_{0} \sim$ \SI{0.9}{\nano\farad\per\meter}, $C_{1} \sim$ \SI{1.1}{\nano\farad\per\meter}. The scaling current was chosen to be \SI{50}{\milli\ampere}, which could be expected from a NbTiN coating deposited on an insulating layer with a rough metallic base (Color figure online).}
    \label{fig:repeatingUnit}	
\end{figure}

Dispersion engineering is achieved in RGWG by employing periodic widening of the ridge as shown in Fig. \ref{fig:repeatingUnit}. The impedance of the loaded and unloaded sections is estimated using $Z=\sqrt{(L_g+gL_{k0})/C}$, where $C$ is the capacitance and $g$ is a geometric factor accounting for the SC coating of both parallel surfaces of the RGWG. More accurate impedance values were extracted from HFSS simulations, these were within $\sim 5\%$ of the analytic solutions. The impedance of the perturbations was $\sim 20\%$ lower than the main line, which introduced a weak dispersion feature near the $f_{per}$ that becomes stronger as the number of repeating units is increased. As a result, a travelling wave structure spanning $450\lambda_{per}$ was investigated. Such an electrically long structure can be simulated directly in HFSS to show the stopband structure, however, HFSS is unable to take into account the current dependency of the KI nor does it support harmonic balance simulations to evaluate mixing. Keysight's ADS\footnote{https://www.keysight.com/us/en/products/software/pathwave-design-software/pathwave-advanced-design-system.html} is an industry standard amplifier design tool that is able to address these aspects of the paramp design process.

\begin{figure}
    \centering
    \begin{subfigure}[b]{0.49\textwidth}
        \centering
        \includegraphics[width=\textwidth]{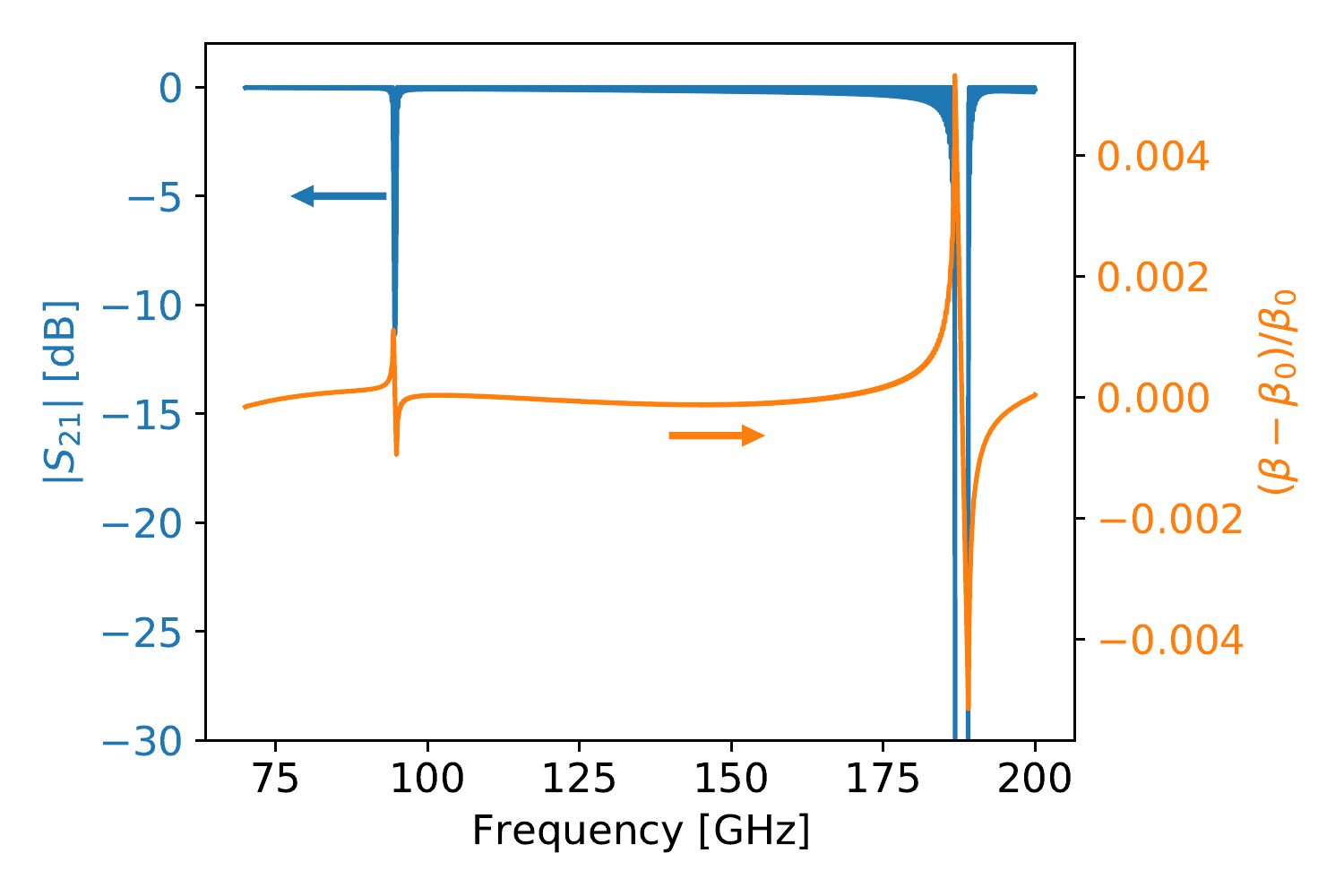}
    \end{subfigure}
    \hfill
    \begin{subfigure}[b]{0.49\textwidth}
        \centering
        \includegraphics[width=\textwidth]{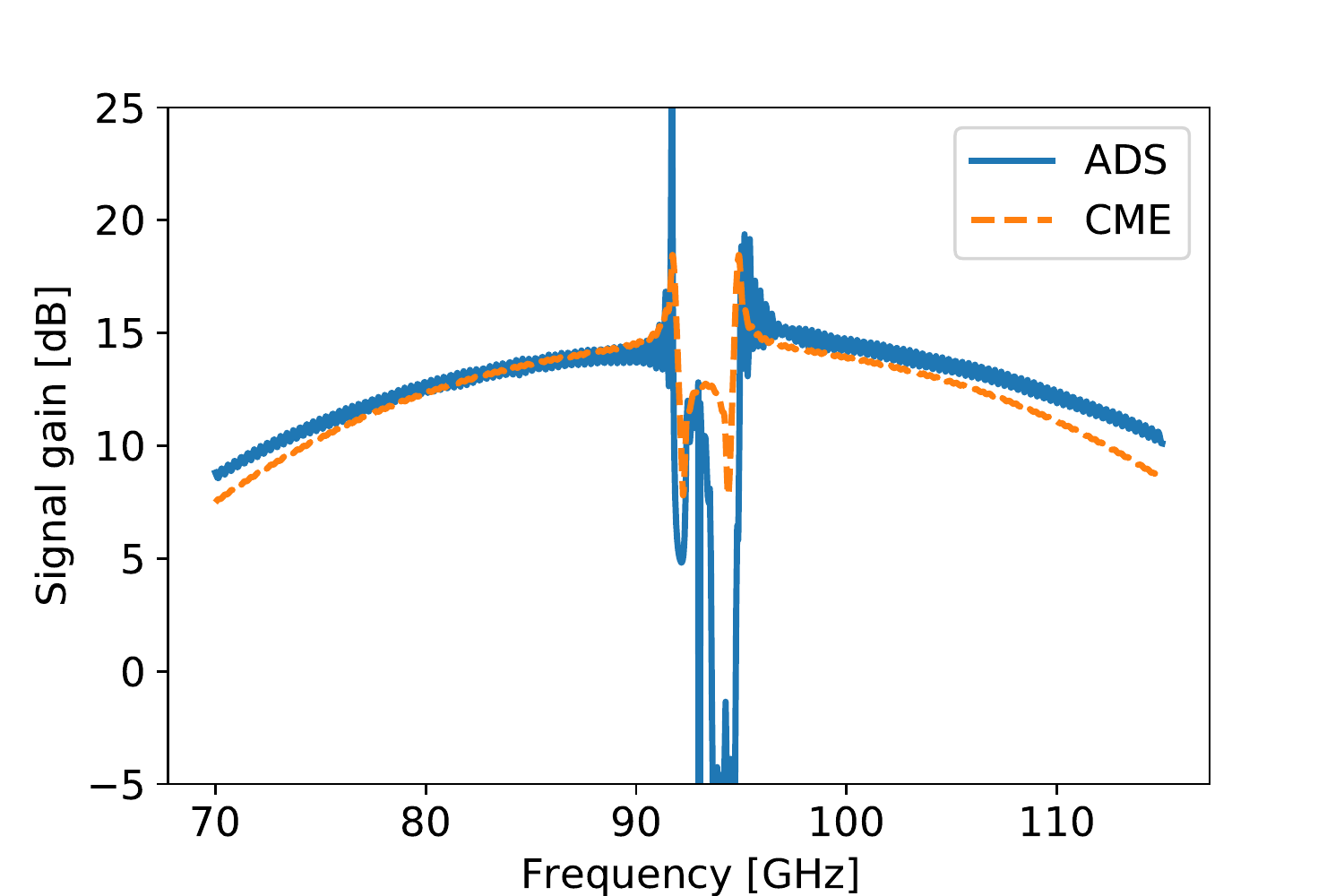}
    \end{subfigure}
	\caption{Left: Transmission coefficient and effective dispersion from a simulation of 900 $\lambda/2$ cells in ADS, corresponding to a physical propagation length of \SI{25}{\centi\meter}. The first 2 stopbands at $f_{per} \sim $ \SI{95}{\giga\hertz} and $f_{per} \sim$ \SI{190}{\giga\hertz} are shown. The unperturbed dispersion $\beta$ was found by fitting a second order polynomial to the perturbed data. Right: The simulated signal gain from the HB simulator of ADS. The -9 dBm pump tone is positioned at \SI{93}{\giga\hertz}, while the signal is at -80 dBm. The CME solution \cite{Chaudhuri2015} uses the dispersion properties extracted from the 900 cell LSSP simulation to calculate the phase mismatch between the 3 involved tones. The calculated CME gain matched the simulated ADS gain at $I/I_*\sim0.19$. (Color figure online).}
    \label{fig:TW-KIPresults}	
\end{figure}

An equivalent circuit model of the nonlinear RGWG was created, which is described in Fig. \ref{fig:repeatingUnit}. This circuit takes into account the nonlinear inductance via a general symbolically defined device (SDD) component, which requires a voltage definition in the form $V = -\frac{d}{dt}L_k(I)I$. These repeating unit cells were nested within subcircuits to make up a total of $N=900$ cells. A large signal S-parameter (LSSP) simulation within ADS was used to generate the nonlinear S-parameters and dispersion properties as shown in Fig. \ref{fig:TW-KIPresults}. The dispersion of the TWPA was calculated via $\beta = -\text{unwrap}[\text{arg}(S_{21})]/L$, where $L$ is the device length.

The parametric gain of the dispersion engineered RGWG can be estimated using the simulated dispersion data and the analytic solutions from the coupled mode equations (CME) \cite{Chaudhuri2015} that describe the amplitudes of the propagating waves. The CME solutions used only take into account the amplitude of the first pump harmonic and the associated mixing products but do not take into account the transmission of the travelling-wave structure. These CMEs can be extended to include higher order pump harmonics \cite{Chaudhuri2015, Dixon2020}, which provide a more accurate gain description. Alternatively, the parametric gain can be simulated \cite{Sweetnam2022} using the harmonic balance (HB) simulator of ADS with the repeating circuit shown in Fig. \ref{fig:repeatingUnit}, which accurately describes the gain of such a paramp to a chosen number of input tone harmonics. The simulated gain of the \SI{25}{\centi\meter} RGWG paramp and is shown in Fig. \ref{fig:TW-KIPresults}, which shows a \SI{10}{\decibel} signal gain over the whole of W-band. This was achieved by sweeping the pump tone frequency and input power close to the first stop band generated by the periodic structure such that the phase mismatch $\Delta\beta$ is minimised and maximum gain is achieved. 

The simulated gain shows promise that the fabricated RGWG KI-TWPA could provide sufficient gain as a front end amplifier. A logical prototype could be a series resonator of similar dimensions which would confirm the feasibility of the high KI RGWG at W-band and would allow the characterisation of the scaling current. The designed RGWG had a conservative conductor spacing of $h = $ \SI{5}{\micro\meter} and a \SI{30}{\nano\meter} NbTiN film coating, which resulted in a KI fraction of $\alpha \sim$ \SI{60}{\percent}. By reducing $h$ to \SI{1}{\micro\meter}, via a shorter deposition of the spacer layer, the value of $\alpha$ may reach $>90\%$, which would significantly reduce the required TWPA length.

The main practical challenges of the RGWG TWPA include the fabrication of the long structure, and the matching of the \SI{6}{\ohm} RGWG impedance to the chosen feedlines. The structure could be fabricated on a 4 inch wafer by reducing the groove width $a$ to \SI{700}{\micro\meter} and meandering the RGWG with a single row of nails between meanders to provide isolation between parallel lines. Transitioning to W band rectangular waveguide could be achieved via a slow tapering of the height of the ridge leading to the waveguide flange. In addition, the ridge could be reduced in width to $<$ \SI{50}{\micro\meter} which would increase the impedance to $>$ \SI{50}{\ohm} and ease the impedance mismatch. The use of an impedance taper would unavoidably cause significant gain ripples. However, such a tapered structure should still be able to demonstrate wideband parametric gain, confirming its' feasibility and paving the way for further development.

\section{Conclusion}

This work proposed a low loss structure for high frequency KI-TWPAs in the form of RGWG. This structure is scaleable to frequencies of a few \SI{100}{\giga\hertz} with feature sizes in the hundreds of \si{\micro\meter} range, meaning that it can be fabricated with high speed metal milling techniques or silicon micromachining and coated in a superconducting film. It was shown via superconducting HFSS simulations that this structure can have over an octave of bandwidth for single mode quasi-TEM transmission, which would be ultimately limited by external factors such as the transition to the chosen waveguide feeds. It was shown via ADS HB simulations that a NbTiN coated RGWG can be the basis of a TWPA and provide over \SI{10}{\decibel} of parametric gain over the whole of W-band (75--110 GHz) with a physical length of \SI{25}{\centi\meter}. This paper sets out the design methods and simulation pipeline for such KI RGWGs and their TWPA counterparts while including an accessible method to simulate the expected parametric gain.

\begin{acknowledgements}
This work was supported by STFC grant ST/R504956/1. D. Banys is supported by a STFC PhD studentship.
\end{acknowledgements}

\pagebreak
\bibliographystyle{aipnum4-1}
\bibliography{references}

\begin{thebibliography}{26}%
\makeatletter
\providecommand \@ifxundefined [1]{%
 \@ifx{#1\undefined}
}%
\providecommand \@ifnum [1]{%
 \ifnum #1\expandafter \@firstoftwo
 \else \expandafter \@secondoftwo
 \fi
}%
\providecommand \@ifx [1]{%
 \ifx #1\expandafter \@firstoftwo
 \else \expandafter \@secondoftwo
 \fi
}%
\providecommand \natexlab [1]{#1}%
\providecommand \enquote  [1]{``#1''}%
\providecommand \bibnamefont  [1]{#1}%
\providecommand \bibfnamefont [1]{#1}%
\providecommand \citenamefont [1]{#1}%
\providecommand \href@noop [0]{\@secondoftwo}%
\providecommand \href [0]{\begingroup \@sanitize@url \@href}%
\providecommand \@href[1]{\@@startlink{#1}\@@href}%
\providecommand \@@href[1]{\endgroup#1\@@endlink}%
\providecommand \@sanitize@url [0]{\catcode `\\12\catcode `\$12\catcode
  `\&12\catcode `\#12\catcode `\^12\catcode `\_12\catcode `\%12\relax}%
\providecommand \@@startlink[1]{}%
\providecommand \@@endlink[0]{}%
\providecommand \url  [0]{\begingroup\@sanitize@url \@url }%
\providecommand \@url [1]{\endgroup\@href {#1}{\urlprefix }}%
\providecommand \urlprefix  [0]{URL }%
\providecommand \Eprint [0]{\href }%
\providecommand \doibase [0]{http://dx.doi.org/}%
\providecommand \selectlanguage [0]{\@gobble}%
\providecommand \bibinfo  [0]{\@secondoftwo}%
\providecommand \bibfield  [0]{\@secondoftwo}%
\providecommand \translation [1]{[#1]}%
\providecommand \BibitemOpen [0]{}%
\providecommand \bibitemStop [0]{}%
\providecommand \bibitemNoStop [0]{.\EOS\space}%
\providecommand \EOS [0]{\spacefactor3000\relax}%
\providecommand \BibitemShut  [1]{\csname bibitem#1\endcsname}%
\let\auto@bib@innerbib\@empty
%</preamble>
\bibitem [{\citenamefont {Esposito}\ \emph {et~al.}(2021)\citenamefont
  {Esposito}, \citenamefont {Ranadive}, \citenamefont {Planat},\ and\
  \citenamefont {Roch}}]{Esposito2021}%
  \BibitemOpen
  \bibfield  {author} {\bibinfo {author} {\bibfnamefont {M.}~\bibnamefont
  {Esposito}}, \bibinfo {author} {\bibfnamefont {A.}~\bibnamefont {Ranadive}},
  \bibinfo {author} {\bibfnamefont {L.}~\bibnamefont {Planat}}, \ and\ \bibinfo
  {author} {\bibfnamefont {N.}~\bibnamefont {Roch}},\ }\href {\doibase
  10.1063/5.0064892} {\bibfield  {journal} {\bibinfo  {journal} {Applied
  Physics Letters}\ }\textbf {\bibinfo {volume} {119}},\ \bibinfo {pages}
  {120501} (\bibinfo {year} {2021})},\ \Eprint
  {http://arxiv.org/abs/https://doi.org/10.1063/5.0064892}
  {https://doi.org/10.1063/5.0064892} \BibitemShut {NoStop}%
\bibitem [{\citenamefont {Eom}\ \emph {et~al.}(2012)\citenamefont {Eom},
  \citenamefont {Day}, \citenamefont {LeDuc},\ and\ \citenamefont
  {Zmuidzinas}}]{Eom2012}%
  \BibitemOpen
  \bibfield  {author} {\bibinfo {author} {\bibfnamefont {B.~H.}\ \bibnamefont
  {Eom}}, \bibinfo {author} {\bibfnamefont {P.~K.}\ \bibnamefont {Day}},
  \bibinfo {author} {\bibfnamefont {H.~G.}\ \bibnamefont {LeDuc}}, \ and\
  \bibinfo {author} {\bibfnamefont {J.}~\bibnamefont {Zmuidzinas}},\ }\href
  {\doibase 10.1038/nphys2356} {\bibfield  {journal} {\bibinfo  {journal}
  {Nature Physics}\ }\textbf {\bibinfo {volume} {8}},\ \bibinfo {pages} {623}
  (\bibinfo {year} {2012})}\BibitemShut {NoStop}%
\bibitem [{\citenamefont {Vissers}\ \emph {et~al.}(2016)\citenamefont
  {Vissers}, \citenamefont {Erickson}, \citenamefont {Ku}, \citenamefont
  {Vale}, \citenamefont {Wu}, \citenamefont {Hilton},\ and\ \citenamefont
  {Pappas}}]{Vissers2016}%
  \BibitemOpen
  \bibfield  {author} {\bibinfo {author} {\bibfnamefont {M.~R.}\ \bibnamefont
  {Vissers}}, \bibinfo {author} {\bibfnamefont {R.~P.}\ \bibnamefont
  {Erickson}}, \bibinfo {author} {\bibfnamefont {H.-S.}\ \bibnamefont {Ku}},
  \bibinfo {author} {\bibfnamefont {L.}~\bibnamefont {Vale}}, \bibinfo {author}
  {\bibfnamefont {X.}~\bibnamefont {Wu}}, \bibinfo {author} {\bibfnamefont
  {G.~C.}\ \bibnamefont {Hilton}}, \ and\ \bibinfo {author} {\bibfnamefont
  {D.~P.}\ \bibnamefont {Pappas}},\ }\href {\doibase 10.1063/1.4937922}
  {\bibfield  {journal} {\bibinfo  {journal} {Applied Physics Letters}\
  }\textbf {\bibinfo {volume} {108}},\ \bibinfo {pages} {012601} (\bibinfo
  {year} {2016})},\ \Eprint
  {http://arxiv.org/abs/https://doi.org/10.1063/1.4937922}
  {https://doi.org/10.1063/1.4937922} \BibitemShut {NoStop}%
\bibitem [{\citenamefont {Malnou}\ \emph
  {et~al.}(2021{\natexlab{a}})\citenamefont {Malnou}, \citenamefont {Vissers},
  \citenamefont {Wheeler}, \citenamefont {Aumentado}, \citenamefont {Hubmayr},
  \citenamefont {Ullom},\ and\ \citenamefont {Gao}}]{Malnou3WM}%
  \BibitemOpen
  \bibfield  {author} {\bibinfo {author} {\bibfnamefont {M.}~\bibnamefont
  {Malnou}}, \bibinfo {author} {\bibfnamefont {M.}~\bibnamefont {Vissers}},
  \bibinfo {author} {\bibfnamefont {J.}~\bibnamefont {Wheeler}}, \bibinfo
  {author} {\bibfnamefont {J.}~\bibnamefont {Aumentado}}, \bibinfo {author}
  {\bibfnamefont {J.}~\bibnamefont {Hubmayr}}, \bibinfo {author} {\bibfnamefont
  {J.}~\bibnamefont {Ullom}}, \ and\ \bibinfo {author} {\bibfnamefont
  {J.}~\bibnamefont {Gao}},\ }\href {\doibase 10.1103/PRXQuantum.2.010302}
  {\bibfield  {journal} {\bibinfo  {journal} {PRX Quantum}\ }\textbf {\bibinfo
  {volume} {2}},\ \bibinfo {pages} {010302} (\bibinfo {year}
  {2021}{\natexlab{a}})}\BibitemShut {NoStop}%
\bibitem [{\citenamefont {Beurthey}\ \emph {et~al.}(2020)\citenamefont
  {Beurthey}, \citenamefont {Böhmer}, \citenamefont {Brun}, \citenamefont
  {Caldwell}, \citenamefont {Chevalier}, \citenamefont {Diaconu}, \citenamefont
  {Dvali}, \citenamefont {Freire}, \citenamefont {Garutti}, \citenamefont
  {Gooch}, \citenamefont {Hambarzumjan}, \citenamefont {Heyminck},
  \citenamefont {Hubaut}, \citenamefont {Jochum}, \citenamefont {Karst},
  \citenamefont {Khan}, \citenamefont {Kittlinger}, \citenamefont {Knirck},
  \citenamefont {Kramer}, \citenamefont {Krieger}, \citenamefont {Lasserre},
  \citenamefont {Lee}, \citenamefont {Li}, \citenamefont {Lindner},
  \citenamefont {Majorovits}, \citenamefont {Matysek}, \citenamefont {Martens},
  \citenamefont {Öz}, \citenamefont {Pataguppi}, \citenamefont {Pralavorio},
  \citenamefont {Raffelt}, \citenamefont {Redondo}, \citenamefont {Reimann},
  \citenamefont {Ringwald}, \citenamefont {Roch}, \citenamefont {Saikawa},
  \citenamefont {Schaffran}, \citenamefont {Schmidt}, \citenamefont
  {Schütte-Engel}, \citenamefont {Sedlak}, \citenamefont {Steffen},
  \citenamefont {Shtembari}, \citenamefont {Strandhagen}, \citenamefont
  {Strom},\ and\ \citenamefont {Wieching}}]{Beurthey2020madmax}%
  \BibitemOpen
  \bibfield  {author} {\bibinfo {author} {\bibfnamefont {S.}~\bibnamefont
  {Beurthey}}, \bibinfo {author} {\bibfnamefont {N.}~\bibnamefont {Böhmer}},
  \bibinfo {author} {\bibfnamefont {P.}~\bibnamefont {Brun}}, \bibinfo {author}
  {\bibfnamefont {A.}~\bibnamefont {Caldwell}}, \bibinfo {author}
  {\bibfnamefont {L.}~\bibnamefont {Chevalier}}, \bibinfo {author}
  {\bibfnamefont {C.}~\bibnamefont {Diaconu}}, \bibinfo {author} {\bibfnamefont
  {G.}~\bibnamefont {Dvali}}, \bibinfo {author} {\bibfnamefont
  {P.}~\bibnamefont {Freire}}, \bibinfo {author} {\bibfnamefont
  {E.}~\bibnamefont {Garutti}}, \bibinfo {author} {\bibfnamefont
  {C.}~\bibnamefont {Gooch}}, \bibinfo {author} {\bibfnamefont
  {A.}~\bibnamefont {Hambarzumjan}}, \bibinfo {author} {\bibfnamefont
  {S.}~\bibnamefont {Heyminck}}, \bibinfo {author} {\bibfnamefont
  {F.}~\bibnamefont {Hubaut}}, \bibinfo {author} {\bibfnamefont
  {J.}~\bibnamefont {Jochum}}, \bibinfo {author} {\bibfnamefont
  {P.}~\bibnamefont {Karst}}, \bibinfo {author} {\bibfnamefont
  {S.}~\bibnamefont {Khan}}, \bibinfo {author} {\bibfnamefont {D.}~\bibnamefont
  {Kittlinger}}, \bibinfo {author} {\bibfnamefont {S.}~\bibnamefont {Knirck}},
  \bibinfo {author} {\bibfnamefont {M.}~\bibnamefont {Kramer}}, \bibinfo
  {author} {\bibfnamefont {C.}~\bibnamefont {Krieger}}, \bibinfo {author}
  {\bibfnamefont {T.}~\bibnamefont {Lasserre}}, \bibinfo {author}
  {\bibfnamefont {C.}~\bibnamefont {Lee}}, \bibinfo {author} {\bibfnamefont
  {X.}~\bibnamefont {Li}}, \bibinfo {author} {\bibfnamefont {A.}~\bibnamefont
  {Lindner}}, \bibinfo {author} {\bibfnamefont {B.}~\bibnamefont {Majorovits}},
  \bibinfo {author} {\bibfnamefont {M.}~\bibnamefont {Matysek}}, \bibinfo
  {author} {\bibfnamefont {S.}~\bibnamefont {Martens}}, \bibinfo {author}
  {\bibfnamefont {E.}~\bibnamefont {Öz}}, \bibinfo {author} {\bibfnamefont
  {P.}~\bibnamefont {Pataguppi}}, \bibinfo {author} {\bibfnamefont
  {P.}~\bibnamefont {Pralavorio}}, \bibinfo {author} {\bibfnamefont
  {G.}~\bibnamefont {Raffelt}}, \bibinfo {author} {\bibfnamefont
  {J.}~\bibnamefont {Redondo}}, \bibinfo {author} {\bibfnamefont
  {O.}~\bibnamefont {Reimann}}, \bibinfo {author} {\bibfnamefont
  {A.}~\bibnamefont {Ringwald}}, \bibinfo {author} {\bibfnamefont
  {N.}~\bibnamefont {Roch}}, \bibinfo {author} {\bibfnamefont {K.}~\bibnamefont
  {Saikawa}}, \bibinfo {author} {\bibfnamefont {J.}~\bibnamefont {Schaffran}},
  \bibinfo {author} {\bibfnamefont {A.}~\bibnamefont {Schmidt}}, \bibinfo
  {author} {\bibfnamefont {J.}~\bibnamefont {Schütte-Engel}}, \bibinfo
  {author} {\bibfnamefont {A.}~\bibnamefont {Sedlak}}, \bibinfo {author}
  {\bibfnamefont {F.}~\bibnamefont {Steffen}}, \bibinfo {author} {\bibfnamefont
  {L.}~\bibnamefont {Shtembari}}, \bibinfo {author} {\bibfnamefont
  {C.}~\bibnamefont {Strandhagen}}, \bibinfo {author} {\bibfnamefont
  {D.}~\bibnamefont {Strom}}, \ and\ \bibinfo {author} {\bibfnamefont
  {G.}~\bibnamefont {Wieching}},\ }\href@noop {} {\enquote {\bibinfo {title}
  {Madmax status report},}\ } (\bibinfo {year} {2020}),\ \Eprint
  {http://arxiv.org/abs/2003.10894} {arXiv:2003.10894 [physics.ins-det]}
  \BibitemShut {NoStop}%
\bibitem [{\citenamefont {Faramarzi}\ \emph {et~al.}(2021)\citenamefont
  {Faramarzi}, \citenamefont {Day}, \citenamefont {Glasby}, \citenamefont
  {Sypkens}, \citenamefont {Colangelo}, \citenamefont {Chamberlin},
  \citenamefont {Mirhosseini}, \citenamefont {Schmidt}, \citenamefont
  {Berggren},\ and\ \citenamefont {Mauskopf}}]{Fermarzi2021}%
  \BibitemOpen
  \bibfield  {author} {\bibinfo {author} {\bibfnamefont {F.}~\bibnamefont
  {Faramarzi}}, \bibinfo {author} {\bibfnamefont {P.}~\bibnamefont {Day}},
  \bibinfo {author} {\bibfnamefont {J.}~\bibnamefont {Glasby}}, \bibinfo
  {author} {\bibfnamefont {S.}~\bibnamefont {Sypkens}}, \bibinfo {author}
  {\bibfnamefont {M.}~\bibnamefont {Colangelo}}, \bibinfo {author}
  {\bibfnamefont {R.}~\bibnamefont {Chamberlin}}, \bibinfo {author}
  {\bibfnamefont {M.}~\bibnamefont {Mirhosseini}}, \bibinfo {author}
  {\bibfnamefont {K.}~\bibnamefont {Schmidt}}, \bibinfo {author} {\bibfnamefont
  {K.~K.}\ \bibnamefont {Berggren}}, \ and\ \bibinfo {author} {\bibfnamefont
  {P.}~\bibnamefont {Mauskopf}},\ }\href {\doibase 10.1109/tasc.2021.3065304}
  {\bibfield  {journal} {\bibinfo  {journal} {IEEE Transactions on Applied
  Superconductivity}\ }\textbf {\bibinfo {volume} {31}},\ \bibinfo {pages}
  {1–5} (\bibinfo {year} {2021})}\BibitemShut {NoStop}%
\bibitem [{\citenamefont {Malnou}\ \emph
  {et~al.}(2021{\natexlab{b}})\citenamefont {Malnou}, \citenamefont
  {Aumentado}, \citenamefont {Vissers}, \citenamefont {Wheeler}, \citenamefont
  {Hubmayr}, \citenamefont {Ullom},\ and\ \citenamefont {Gao}}]{Malnou2021_4K}%
  \BibitemOpen
  \bibfield  {author} {\bibinfo {author} {\bibfnamefont {M.}~\bibnamefont
  {Malnou}}, \bibinfo {author} {\bibfnamefont {J.}~\bibnamefont {Aumentado}},
  \bibinfo {author} {\bibfnamefont {M.~R.}\ \bibnamefont {Vissers}}, \bibinfo
  {author} {\bibfnamefont {J.~D.}\ \bibnamefont {Wheeler}}, \bibinfo {author}
  {\bibfnamefont {J.}~\bibnamefont {Hubmayr}}, \bibinfo {author} {\bibfnamefont
  {J.~N.}\ \bibnamefont {Ullom}}, \ and\ \bibinfo {author} {\bibfnamefont
  {J.}~\bibnamefont {Gao}},\ }\href@noop {} {\enquote {\bibinfo {title}
  {Performance of a kinetic-inductance traveling-wave parametric amplifier at 4
  kelvin: Toward an alternative to semiconductor amplifiers},}\ } (\bibinfo
  {year} {2021}{\natexlab{b}}),\ \Eprint {http://arxiv.org/abs/2110.08142}
  {arXiv:2110.08142 [quant-ph]} \BibitemShut {NoStop}%
\bibitem [{\citenamefont {Banys}\ \emph {et~al.}(2020)\citenamefont {Banys},
  \citenamefont {McCulloch}, \citenamefont {Azzoni}, \citenamefont {Cooper},
  \citenamefont {May}, \citenamefont {Melhuish}, \citenamefont {Piccirillo},\
  and\ \citenamefont {Wenninger}}]{Banys2020}%
  \BibitemOpen
  \bibfield  {author} {\bibinfo {author} {\bibfnamefont {D.}~\bibnamefont
  {Banys}}, \bibinfo {author} {\bibfnamefont {M.~A.}\ \bibnamefont
  {McCulloch}}, \bibinfo {author} {\bibfnamefont {S.}~\bibnamefont {Azzoni}},
  \bibinfo {author} {\bibfnamefont {B.}~\bibnamefont {Cooper}}, \bibinfo
  {author} {\bibfnamefont {A.~J.}\ \bibnamefont {May}}, \bibinfo {author}
  {\bibfnamefont {S.~J.}\ \bibnamefont {Melhuish}}, \bibinfo {author}
  {\bibfnamefont {L.}~\bibnamefont {Piccirillo}}, \ and\ \bibinfo {author}
  {\bibfnamefont {J.}~\bibnamefont {Wenninger}},\ }\href {\doibase
  10.1007/s10909-020-02439-w} {\bibfield  {journal} {\bibinfo  {journal}
  {Journal of Low Temperature Physics}\ }\textbf {\bibinfo {volume} {200}},\
  \bibinfo {pages} {295} (\bibinfo {year} {2020})}\BibitemShut {NoStop}%
\bibitem [{\citenamefont {Anferov}\ \emph {et~al.}(2020)\citenamefont
  {Anferov}, \citenamefont {Suleymanzade}, \citenamefont {Oriani},
  \citenamefont {Simon},\ and\ \citenamefont {Schuster}}]{Anferov2020}%
  \BibitemOpen
  \bibfield  {author} {\bibinfo {author} {\bibfnamefont {A.}~\bibnamefont
  {Anferov}}, \bibinfo {author} {\bibfnamefont {A.}~\bibnamefont
  {Suleymanzade}}, \bibinfo {author} {\bibfnamefont {A.}~\bibnamefont
  {Oriani}}, \bibinfo {author} {\bibfnamefont {J.}~\bibnamefont {Simon}}, \
  and\ \bibinfo {author} {\bibfnamefont {D.~I.}\ \bibnamefont {Schuster}},\
  }\href {\doibase 10.1103/PhysRevApplied.13.024056} {\bibfield  {journal}
  {\bibinfo  {journal} {Phys. Rev. Applied}\ }\textbf {\bibinfo {volume}
  {13}},\ \bibinfo {pages} {024056} (\bibinfo {year} {2020})}\BibitemShut
  {NoStop}%
\bibitem [{\citenamefont {Shu}\ \emph {et~al.}(2021)\citenamefont {Shu},
  \citenamefont {Klimovich}, \citenamefont {Eom}, \citenamefont {Beyer},
  \citenamefont {Thakur}, \citenamefont {Leduc},\ and\ \citenamefont
  {Day}}]{Shu2021}%
  \BibitemOpen
  \bibfield  {author} {\bibinfo {author} {\bibfnamefont {S.}~\bibnamefont
  {Shu}}, \bibinfo {author} {\bibfnamefont {N.}~\bibnamefont {Klimovich}},
  \bibinfo {author} {\bibfnamefont {B.~H.}\ \bibnamefont {Eom}}, \bibinfo
  {author} {\bibfnamefont {A.~D.}\ \bibnamefont {Beyer}}, \bibinfo {author}
  {\bibfnamefont {R.~B.}\ \bibnamefont {Thakur}}, \bibinfo {author}
  {\bibfnamefont {H.~G.}\ \bibnamefont {Leduc}}, \ and\ \bibinfo {author}
  {\bibfnamefont {P.~K.}\ \bibnamefont {Day}},\ }\href {\doibase
  10.1103/PhysRevResearch.3.023184} {\bibfield  {journal} {\bibinfo  {journal}
  {Phys. Rev. Research}\ }\textbf {\bibinfo {volume} {3}},\ \bibinfo {pages}
  {023184} (\bibinfo {year} {2021})}\BibitemShut {NoStop}%
\bibitem [{\citenamefont {Kildal}\ \emph {et~al.}(2011)\citenamefont {Kildal},
  \citenamefont {Zaman}, \citenamefont {Rajo-Iglesias}, \citenamefont
  {Alfonso},\ and\ \citenamefont {Valero-Nogueira}}]{Kildal2011}%
  \BibitemOpen
  \bibfield  {author} {\bibinfo {author} {\bibfnamefont {P.-S.}\ \bibnamefont
  {Kildal}}, \bibinfo {author} {\bibfnamefont {A.}~\bibnamefont {Zaman}},
  \bibinfo {author} {\bibfnamefont {E.}~\bibnamefont {Rajo-Iglesias}}, \bibinfo
  {author} {\bibfnamefont {E.}~\bibnamefont {Alfonso}}, \ and\ \bibinfo
  {author} {\bibfnamefont {A.}~\bibnamefont {Valero-Nogueira}},\ }\href
  {https://digital-library.theiet.org/content/journals/10.1049/iet-map.2010.0089}
  {\bibfield  {journal} {\bibinfo  {journal} {IET Microwaves, Antennas \&
  Propagation}\ }\textbf {\bibinfo {volume} {5}},\ \bibinfo {pages} {262}
  (\bibinfo {year} {2011})}\BibitemShut {NoStop}%
\bibitem [{\citenamefont {Rahiminejad}\ \emph {et~al.}(2012)\citenamefont
  {Rahiminejad}, \citenamefont {Zaman}, \citenamefont {Pucci}, \citenamefont
  {Raza}, \citenamefont {Vassilev}, \citenamefont {Haasl}, \citenamefont
  {Lundgren}, \citenamefont {Kildal},\ and\ \citenamefont
  {Enoksson}}]{Rahiminejad2012}%
  \BibitemOpen
  \bibfield  {author} {\bibinfo {author} {\bibfnamefont {S.}~\bibnamefont
  {Rahiminejad}}, \bibinfo {author} {\bibfnamefont {A.}~\bibnamefont {Zaman}},
  \bibinfo {author} {\bibfnamefont {E.}~\bibnamefont {Pucci}}, \bibinfo
  {author} {\bibfnamefont {H.}~\bibnamefont {Raza}}, \bibinfo {author}
  {\bibfnamefont {V.}~\bibnamefont {Vassilev}}, \bibinfo {author}
  {\bibfnamefont {S.}~\bibnamefont {Haasl}}, \bibinfo {author} {\bibfnamefont
  {P.}~\bibnamefont {Lundgren}}, \bibinfo {author} {\bibfnamefont {P.-S.}\
  \bibnamefont {Kildal}}, \ and\ \bibinfo {author} {\bibfnamefont
  {P.}~\bibnamefont {Enoksson}},\ }\href {\doibase
  https://doi.org/10.1016/j.sna.2012.02.036} {\bibfield  {journal} {\bibinfo
  {journal} {Sensors and Actuators A: Physical}\ }\textbf {\bibinfo {volume}
  {186}},\ \bibinfo {pages} {264} (\bibinfo {year} {2012})},\ \bibinfo {note}
  {selected Papers presented at Eurosensors XXV}\BibitemShut {NoStop}%
\bibitem [{\citenamefont {Belitsky}\ \emph {et~al.}(2006)\citenamefont
  {Belitsky}, \citenamefont {Risacher}, \citenamefont {Pantaleev},\ and\
  \citenamefont {Vassilev}}]{Belitsky2006}%
  \BibitemOpen
  \bibfield  {author} {\bibinfo {author} {\bibfnamefont {V.}~\bibnamefont
  {Belitsky}}, \bibinfo {author} {\bibfnamefont {C.}~\bibnamefont {Risacher}},
  \bibinfo {author} {\bibfnamefont {M.}~\bibnamefont {Pantaleev}}, \ and\
  \bibinfo {author} {\bibfnamefont {V.}~\bibnamefont {Vassilev}},\ }\href
  {\doibase 10.1007/s10762-006-9116-5} {\bibfield  {journal} {\bibinfo
  {journal} {International Journal of Infrared and Millimeter Waves}\ }\textbf
  {\bibinfo {volume} {27}},\ \bibinfo {pages} {809} (\bibinfo {year}
  {2006})}\BibitemShut {NoStop}%
\bibitem [{\citenamefont {Zimmermann}\ \emph {et~al.}(1991)\citenamefont
  {Zimmermann}, \citenamefont {Brandt}, \citenamefont {Bauer}, \citenamefont
  {Seider},\ and\ \citenamefont {Genzel}}]{Zimmermann1991}%
  \BibitemOpen
  \bibfield  {author} {\bibinfo {author} {\bibfnamefont {W.}~\bibnamefont
  {Zimmermann}}, \bibinfo {author} {\bibfnamefont {E.}~\bibnamefont {Brandt}},
  \bibinfo {author} {\bibfnamefont {M.}~\bibnamefont {Bauer}}, \bibinfo
  {author} {\bibfnamefont {E.}~\bibnamefont {Seider}}, \ and\ \bibinfo {author}
  {\bibfnamefont {L.}~\bibnamefont {Genzel}},\ }\href {\doibase
  https://doi.org/10.1016/0921-4534(91)90771-P} {\bibfield  {journal} {\bibinfo
   {journal} {Physica C: Superconductivity}\ }\textbf {\bibinfo {volume}
  {183}},\ \bibinfo {pages} {99} (\bibinfo {year} {1991})}\BibitemShut
  {NoStop}%
\bibitem [{\citenamefont {Gurevich}(2017)}]{Gurevich2017}%
  \BibitemOpen
  \bibfield  {author} {\bibinfo {author} {\bibfnamefont {A.}~\bibnamefont
  {Gurevich}},\ }\href {\doibase 10.1088/1361-6668/30/3/034004} {\bibfield
  {journal} {\bibinfo  {journal} {Superconductor Science and Technology}\
  }\textbf {\bibinfo {volume} {30}},\ \bibinfo {pages} {034004} (\bibinfo
  {year} {2017})}\BibitemShut {NoStop}%
\bibitem [{\citenamefont {Semenov}\ \emph {et~al.}(2020)\citenamefont
  {Semenov}, \citenamefont {Devyatov}, \citenamefont {Westig},\ and\
  \citenamefont {Klapwijk}}]{Semenov2020}%
  \BibitemOpen
  \bibfield  {author} {\bibinfo {author} {\bibfnamefont {A.}~\bibnamefont
  {Semenov}}, \bibinfo {author} {\bibfnamefont {I.}~\bibnamefont {Devyatov}},
  \bibinfo {author} {\bibfnamefont {M.}~\bibnamefont {Westig}}, \ and\ \bibinfo
  {author} {\bibfnamefont {T.}~\bibnamefont {Klapwijk}},\ }\href {\doibase
  10.1103/PhysRevApplied.13.024079} {\bibfield  {journal} {\bibinfo  {journal}
  {Phys. Rev. Applied}\ }\textbf {\bibinfo {volume} {13}},\ \bibinfo {pages}
  {024079} (\bibinfo {year} {2020})}\BibitemShut {NoStop}%
\bibitem [{\citenamefont {Zolochevskii}(2014)}]{Zolochevskii2014}%
  \BibitemOpen
  \bibfield  {author} {\bibinfo {author} {\bibfnamefont {I.~V.}\ \bibnamefont
  {Zolochevskii}},\ }\href {\doibase 10.1063/1.4900695} {\bibfield  {journal}
  {\bibinfo  {journal} {Low Temperature Physics}\ }\textbf {\bibinfo {volume}
  {40}},\ \bibinfo {pages} {867} (\bibinfo {year} {2014})},\ \Eprint
  {http://arxiv.org/abs/https://doi.org/10.1063/1.4900695}
  {https://doi.org/10.1063/1.4900695} \BibitemShut {NoStop}%
\bibitem [{\citenamefont {Pearl}(1964)}]{Pearl1964}%
  \BibitemOpen
  \bibfield  {author} {\bibinfo {author} {\bibfnamefont {J.}~\bibnamefont
  {Pearl}},\ }\href {\doibase 10.1063/1.1754056} {\bibfield  {journal}
  {\bibinfo  {journal} {Applied Physics Letters}\ }\textbf {\bibinfo {volume}
  {5}},\ \bibinfo {pages} {65} (\bibinfo {year} {1964})},\ \Eprint
  {http://arxiv.org/abs/https://doi.org/10.1063/1.1754056}
  {https://doi.org/10.1063/1.1754056} \BibitemShut {NoStop}%
\bibitem [{\citenamefont {Zmuidzinas}(2012)}]{Zmuidzinas2012}%
  \BibitemOpen
  \bibfield  {author} {\bibinfo {author} {\bibfnamefont {J.}~\bibnamefont
  {Zmuidzinas}},\ }\href {\doibase 10.1146/annurev-conmatphys-020911-125022}
  {\bibfield  {journal} {\bibinfo  {journal} {Annual Review of Condensed Matter
  Physics}\ }\textbf {\bibinfo {volume} {3}},\ \bibinfo {pages} {169} (\bibinfo
  {year} {2012})},\ \Eprint
  {http://arxiv.org/abs/https://doi.org/10.1146/annurev-conmatphys-020911-125022}
  {https://doi.org/10.1146/annurev-conmatphys-020911-125022} \BibitemShut
  {NoStop}%
\bibitem [{\citenamefont {Zhao}\ \emph {et~al.}(2020)\citenamefont {Zhao},
  \citenamefont {Withington}, \citenamefont {Goldie},\ and\ \citenamefont
  {Thomas}}]{Zhao2020}%
  \BibitemOpen
  \bibfield  {author} {\bibinfo {author} {\bibfnamefont {S.}~\bibnamefont
  {Zhao}}, \bibinfo {author} {\bibfnamefont {S.}~\bibnamefont {Withington}},
  \bibinfo {author} {\bibfnamefont {D.~J.}\ \bibnamefont {Goldie}}, \ and\
  \bibinfo {author} {\bibfnamefont {C.~N.}\ \bibnamefont {Thomas}},\ }\href
  {\doibase 10.1007/s10909-019-02306-3} {\bibfield  {journal} {\bibinfo
  {journal} {Journal of Low Temperature Physics}\ }\textbf {\bibinfo {volume}
  {199}},\ \bibinfo {pages} {34} (\bibinfo {year} {2020})}\BibitemShut
  {NoStop}%
\bibitem [{\citenamefont {Erickson}\ and\ \citenamefont
  {Pappas}(2017)}]{Erickson2017}%
  \BibitemOpen
  \bibfield  {author} {\bibinfo {author} {\bibfnamefont {R.~P.}\ \bibnamefont
  {Erickson}}\ and\ \bibinfo {author} {\bibfnamefont {D.~P.}\ \bibnamefont
  {Pappas}},\ }\href {\doibase 10.1103/PhysRevB.95.104506} {\bibfield
  {journal} {\bibinfo  {journal} {Phys. Rev. B}\ }\textbf {\bibinfo {volume}
  {95}},\ \bibinfo {pages} {104506} (\bibinfo {year} {2017})}\BibitemShut
  {NoStop}%
\bibitem [{\citenamefont {Bockstiegel}\ \emph {et~al.}(2014)\citenamefont
  {Bockstiegel}, \citenamefont {Gao}, \citenamefont {Vissers}, \citenamefont
  {Sandberg}, \citenamefont {Chaudhuri}, \citenamefont {Sanders}, \citenamefont
  {Vale}, \citenamefont {Irwin},\ and\ \citenamefont
  {Pappas}}]{Bockstiegel2014}%
  \BibitemOpen
  \bibfield  {author} {\bibinfo {author} {\bibfnamefont {C.}~\bibnamefont
  {Bockstiegel}}, \bibinfo {author} {\bibfnamefont {J.}~\bibnamefont {Gao}},
  \bibinfo {author} {\bibfnamefont {M.~R.}\ \bibnamefont {Vissers}}, \bibinfo
  {author} {\bibfnamefont {M.}~\bibnamefont {Sandberg}}, \bibinfo {author}
  {\bibfnamefont {S.}~\bibnamefont {Chaudhuri}}, \bibinfo {author}
  {\bibfnamefont {A.}~\bibnamefont {Sanders}}, \bibinfo {author} {\bibfnamefont
  {L.~R.}\ \bibnamefont {Vale}}, \bibinfo {author} {\bibfnamefont {K.~D.}\
  \bibnamefont {Irwin}}, \ and\ \bibinfo {author} {\bibfnamefont {D.~P.}\
  \bibnamefont {Pappas}},\ }\href {\doibase 10.1007/s10909-013-1042-z}
  {\bibfield  {journal} {\bibinfo  {journal} {Journal of Low Temperature
  Physics}\ }\textbf {\bibinfo {volume} {176}},\ \bibinfo {pages} {476}
  (\bibinfo {year} {2014})}\BibitemShut {NoStop}%
\bibitem [{\citenamefont {Shan}, \citenamefont {Sekimoto},\ and\ \citenamefont
  {Noguchi}(2016)}]{Shan2016}%
  \BibitemOpen
  \bibfield  {author} {\bibinfo {author} {\bibfnamefont {W.}~\bibnamefont
  {Shan}}, \bibinfo {author} {\bibfnamefont {Y.}~\bibnamefont {Sekimoto}}, \
  and\ \bibinfo {author} {\bibfnamefont {T.}~\bibnamefont {Noguchi}},\ }\href
  {\doibase 10.1109/TASC.2016.2555914} {\bibfield  {journal} {\bibinfo
  {journal} {IEEE Transactions on Applied Superconductivity}\ }\textbf
  {\bibinfo {volume} {26}},\ \bibinfo {pages} {1} (\bibinfo {year}
  {2016})}\BibitemShut {NoStop}%
\bibitem [{\citenamefont {Chaudhuri}, \citenamefont {Gao},\ and\ \citenamefont
  {Irwin}(2015)}]{Chaudhuri2015}%
  \BibitemOpen
  \bibfield  {author} {\bibinfo {author} {\bibfnamefont {S.}~\bibnamefont
  {Chaudhuri}}, \bibinfo {author} {\bibfnamefont {J.}~\bibnamefont {Gao}}, \
  and\ \bibinfo {author} {\bibfnamefont {K.}~\bibnamefont {Irwin}},\ }\href
  {\doibase 10.1109/TASC.2014.2378059} {\bibfield  {journal} {\bibinfo
  {journal} {IEEE Transactions on Applied Superconductivity}\ }\textbf
  {\bibinfo {volume} {25}},\ \bibinfo {pages} {1} (\bibinfo {year}
  {2015})}\BibitemShut {NoStop}%
\bibitem [{\citenamefont {Dixon}\ \emph {et~al.}(2020)\citenamefont {Dixon},
  \citenamefont {Dunstan}, \citenamefont {Long}, \citenamefont {Williams},
  \citenamefont {Meeson},\ and\ \citenamefont {Shelly}}]{Dixon2020}%
  \BibitemOpen
  \bibfield  {author} {\bibinfo {author} {\bibfnamefont {T.}~\bibnamefont
  {Dixon}}, \bibinfo {author} {\bibfnamefont {J.}~\bibnamefont {Dunstan}},
  \bibinfo {author} {\bibfnamefont {G.}~\bibnamefont {Long}}, \bibinfo {author}
  {\bibfnamefont {J.}~\bibnamefont {Williams}}, \bibinfo {author}
  {\bibfnamefont {P.}~\bibnamefont {Meeson}}, \ and\ \bibinfo {author}
  {\bibfnamefont {C.}~\bibnamefont {Shelly}},\ }\href {\doibase
  10.1103/PhysRevApplied.14.034058} {\bibfield  {journal} {\bibinfo  {journal}
  {Phys. Rev. Applied}\ }\textbf {\bibinfo {volume} {14}},\ \bibinfo {pages}
  {034058} (\bibinfo {year} {2020})}\BibitemShut {NoStop}%
\bibitem [{\citenamefont {Sweetnam}\ \emph {et~al.}(2022)\citenamefont
  {Sweetnam}, \citenamefont {Banys}, \citenamefont {Gilles}, \citenamefont
  {McCulloch},\ and\ \citenamefont {Piccirillo}}]{Sweetnam2022}%
  \BibitemOpen
  \bibfield  {author} {\bibinfo {author} {\bibfnamefont {T.}~\bibnamefont
  {Sweetnam}}, \bibinfo {author} {\bibfnamefont {D.}~\bibnamefont {Banys}},
  \bibinfo {author} {\bibfnamefont {V.}~\bibnamefont {Gilles}}, \bibinfo
  {author} {\bibfnamefont {M.}~\bibnamefont {McCulloch}}, \ and\ \bibinfo
  {author} {\bibfnamefont {L.}~\bibnamefont {Piccirillo}},\ }\href {\doibase
  10.1088/1361-6668/ac850b} {\bibfield  {journal} {\bibinfo  {journal}
  {Superconductor Science and Technology}\ } (\bibinfo {year} {2022}),\
  10.1088/1361-6668/ac850b}\BibitemShut {NoStop}%
\end{thebibliography}%

\end{document}